\documentclass[opinion]{sigirforum}
\usepackage{subcaption}
\usepackage{placeins}
\def\pubissue{Vol. 57 No. 2 -- December 2023}

\begin{document}
\title{Rethinking E-Commerce Search}

\authors{
\author[haixun.wang@instacart.com]{Haixun Wang}{Instacart, Inc.}{USA}
\and
\author[taesik.na@instacart.com]{Taesik Na}{Instacart, Inc.}{USA}
}

\maketitle 
\begin{abstract}

E-commerce search and recommendation usually operate on structured data such as product catalogs and taxonomies. However, creating better search and recommendation systems often requires a large variety of unstructured data including customer reviews and articles on the web. Traditionally, the solution has always been converting unstructured data into structured data through information extraction, and conducting search over the structured data. However, this is a  costly approach that often has low quality. In this paper, we envision a solution that does entirely the opposite. Instead of converting unstructured data (web pages, customer reviews, etc) to structured data, we instead convert structured data (product inventory, catalogs, taxonomies, etc) into textual data, which can be easily integrated into the text corpus that trains LLMs. Then, search and recommendation can be performed through a Q/A mechanism through an LLM instead of using traditional information retrieval methods over structured data.

\end{abstract}

\section{Introduction}

\begin{figure*}[t!]
    \centering
    \begin{subfigure}{0.3\textwidth}
        \centering
        \includegraphics[height=3in]{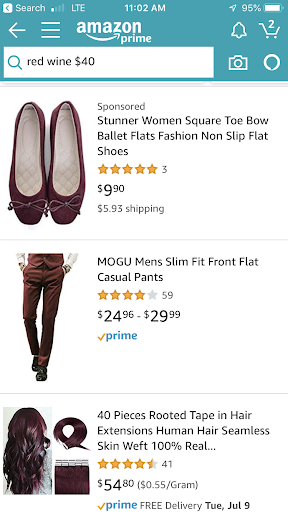}
        \caption{July 7, 2019}
    \end{subfigure}%
    ~ 
    \begin{subfigure}{0.3\textwidth}
        \centering
        \includegraphics[height=3in]{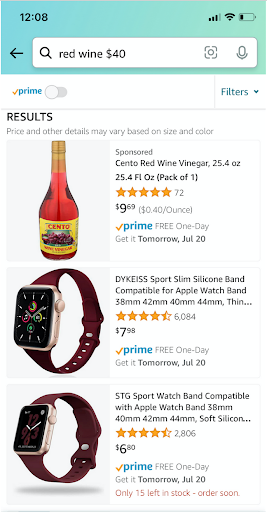}
        \caption{July 19, 2021}
    \end{subfigure}
    ~ 
    \begin{subfigure}{0.3\textwidth}
        \centering
        \includegraphics[height=3in]{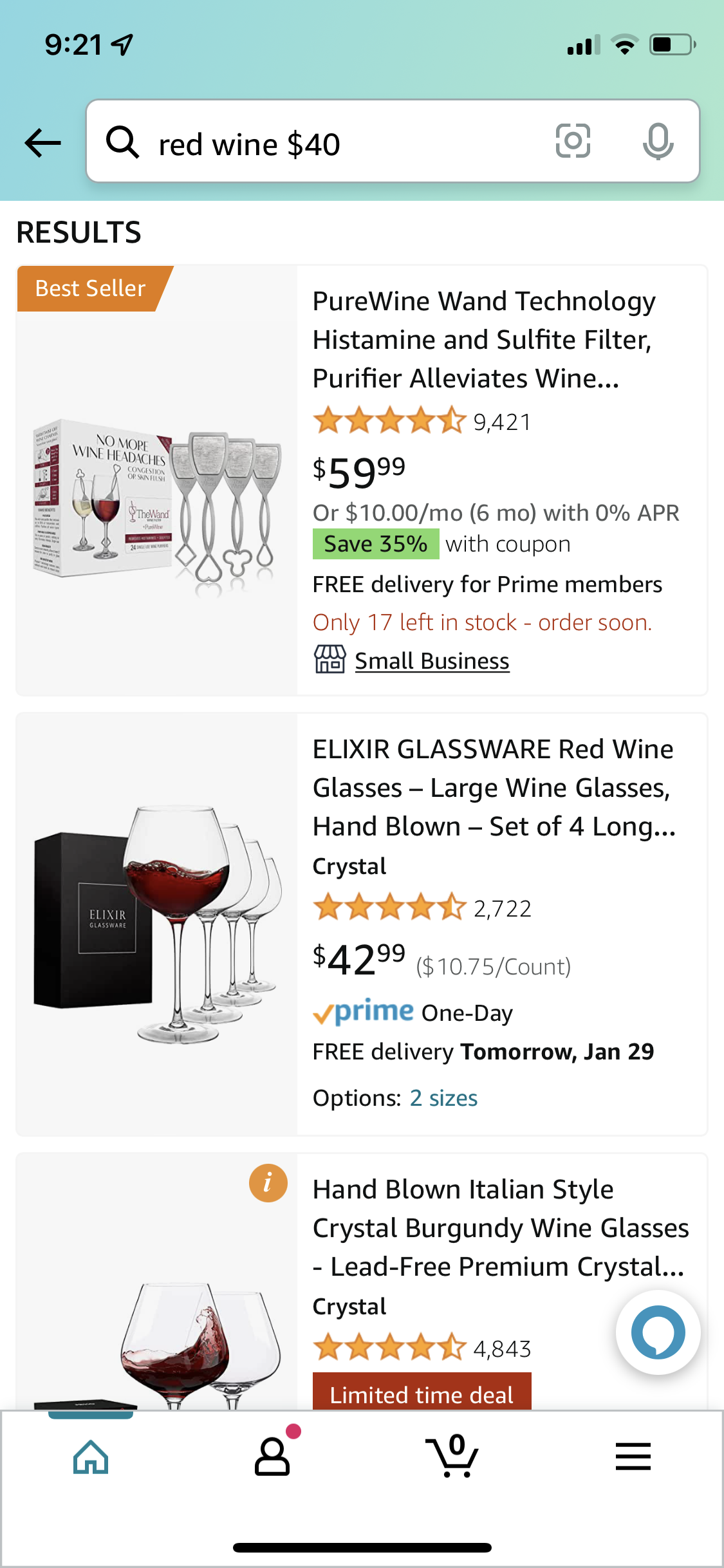}
        \caption{Jan 28, 2023}
    \end{subfigure}    \caption{Results for query ``red wine \$40" on Amazon from 2019 to 2023}
\label{fig:amazon}
\end{figure*}

E-commerce search primarily operates in the ``structured data" domain, with the objective of finding the most relevant products in a product database. While the database contains textual attributes such as product names and descriptions, it is predominantly a structured dataset containing records with attributes such as brand, price, size, weight, and other product-specific features such as gluten-free, screen resolution, etc.

\subsection{Challenges in E-Commerce Search}
E-commerce search presents two fundamental challenges. The first is understanding the user intent behind short search queries, and the second is understanding products in the catalog, which requires world knowledge or a product knowledge base.

\subsubsection{{Query Understanding}}

In classic information retrieval, the challenge of query understanding is largely the challenge of extracting information from the query and map it to database attributes such as brand, product name, and other features such as organic.

But even for a search query as simple as ``red wine \$40", most e-commerce search engines fail to understand the intent behind the search. As shown in Fig~\ref{fig:amazon}, over the years, Amazon returned a variety of products other than red wine. A red-colored pair of shoes, because the color of the shoes or pants turned out to be “wine red.” The system also failed to understand \$40 is an approximate price constraint, and returned pants of waist size 40. These are well-known issues for query understanding, but in practice, there are numerous edge cases that require specialized algorithms and models. As a result, query understanding remained largely an unsolved problem in industrial e-commerce search systems.

\subsubsection{{Document (Product) Understanding}}
Each product in the product database (catalog) has attributes such as name, description, price, and so on. It is not easy to find the most relevant products for a given  query through term-based IR methods. Recent development in embedding-based retrieval~\citep{zhang2020retrieval, huang2020embedding, na2022embedding} has improved semantic matching, but it is still difficult to answer queries such as ``healthy snacks for kids,'' or ``alternatives to ice cream.''

Clearly, E-commerce search is a knowledge-intensive task. Providing customers with recommendations, ideas, and inspiration requires knowledge outside of the product database. E-commerce companies such as Amazon and Baidu invest heavily in developing product knowledge graphs with the goal of learning everything there is to know about their products. 

Unfortunately, building such product knowledge graphs has had limited success to date. At least two obstacles exist. First, modeling such knowledge explicitly in a product knowledge graph necessitates the creation of a very complex schema or ontology, which makes accessing such knowledge challenging because algorithms must navigate the complex schema or ontology. Second, it is difficult to develop a general-purpose algorithm that can convert unstructured data about any product into a structured format, because each product type may require the development of algorithms that identify patterns that are specific to that product type.

\subsection{A New Vision for E-Commerce Search}

\begin{figure}[!p]
	\centering
	\includegraphics[width=0.9\textwidth]{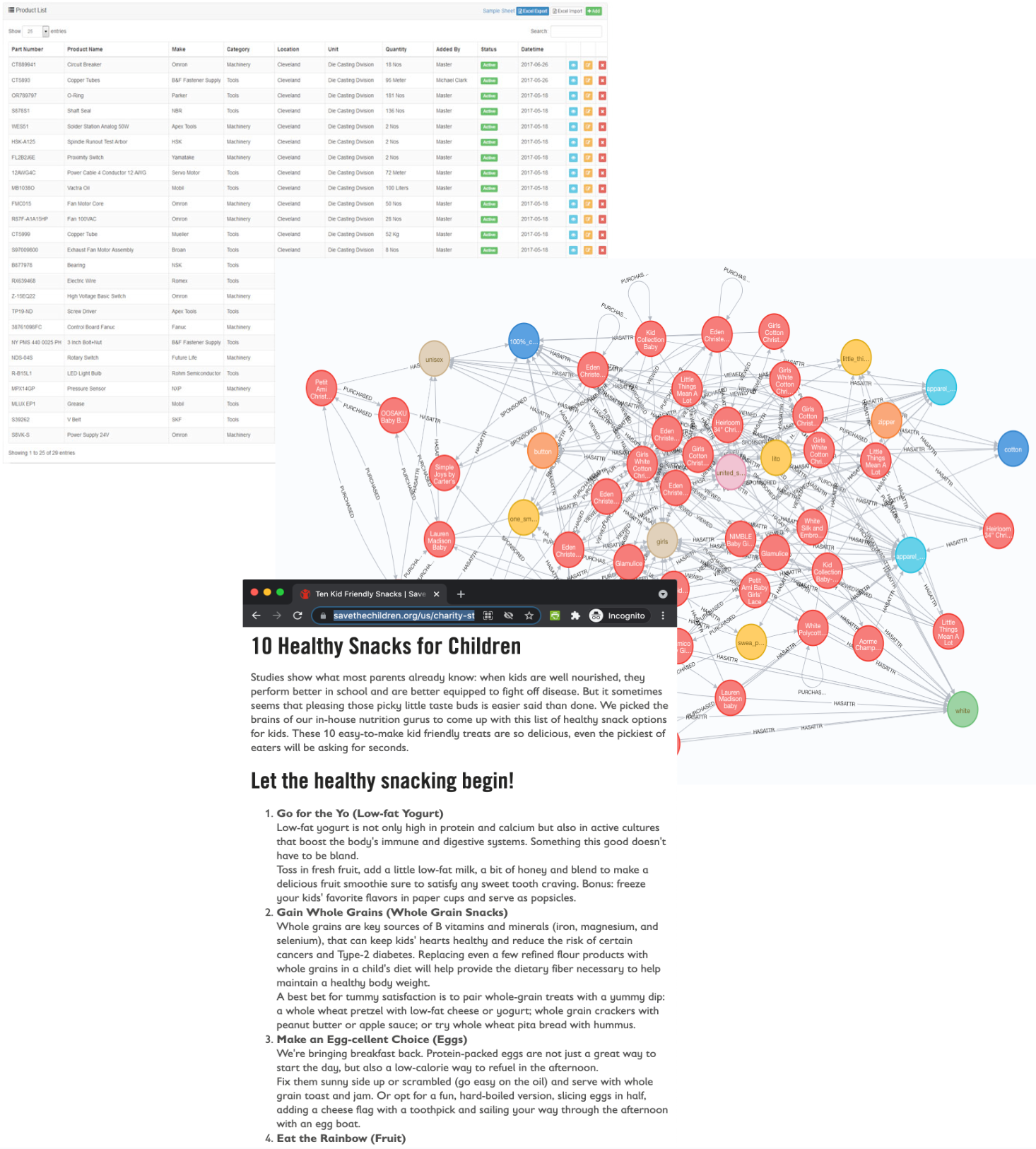}
	\caption{The success of e-commerce search requires the integration of structured, semi-structured, and unstructured data.}
	\label{fig:integration}
\end{figure}

E-commerce search is a knowledge-intensive task. As shown in Fig~\ref{fig:integration}, its success relies on the seamless integration of structured data (e.g., the product catalog), semi-structured data (e.g., taxonomy and ontology), and unstructured data (e.g., customer reviews, web articles, etc.) 

In fact, this problem is more general than e-commerce search, as it applies to any task that involves querying against a database, while some critical knowledge for answering such queries resides in data of heterogeneous types outside of the database.

For such problems, the traditional approach is to collect all the relevant data and convert them into a structured form such that they can be indexed, retrieved, and ranked for  search. The challenge is that the data could be massive, and the cost of information extraction  could be prohibitively expensive.

In this paper, we envision a solution that is the direct opposite of what the e-commerce industry has done in recent decades. Rather than converting all  information into a structured form and then conducting search over a database, we convert all structured and semi-structured data into text and then answer queries through a large language model (LLM) trained over the text. 

The following is the rationale for our new approach. First, answering questions like ``healthy snacks for kids" or ``alternatives to ice creams" requires world knowledge, which is difficult to capture in product knowledge graphs but readily available in pre-trained LLMs like GPT-3~\citep{gpt3}. Second, LLMs provide a text-to-text interface~\citep{t5}, eliminating the need for a separate query understanding system. Third, as we will see, while converting structured data to text is not trivial, the relevant structured data for any specific task such as e-commerce search  is not only limited in size and variety, but also well understood due to the presence of a schema. Therefore, we can focus on developing algorithms to properly convert structured data into additional text for training LLMs.

\subsection{Paper Organization}
This position paper envisions a new solution for E-commerce search. Section ~\ref{sec:motivation} describes the motivation and design of a new architecture for information retrieval.
Section ~\ref{sec:ids} discusses how to better represent the database IDs.
In Section ~\ref{sec:data}, we propose methods to convert structured and semi-structured data to annotated text.
Section ~\ref{sec:system} discusses fine-tuning strategies for LLMs and inference time use cases. Section ~\ref{sec:research}, discusses research direction. We discuss related work in Section ~\ref{sec:related} before we conclude in Section ~\ref{sec:conclusion}.

\pagebreak\section{A Database and LLM Powered IR System}
\label{sec:motivation}

\begin{figure}[!h]
  \centering
  \includegraphics[width=12cm]{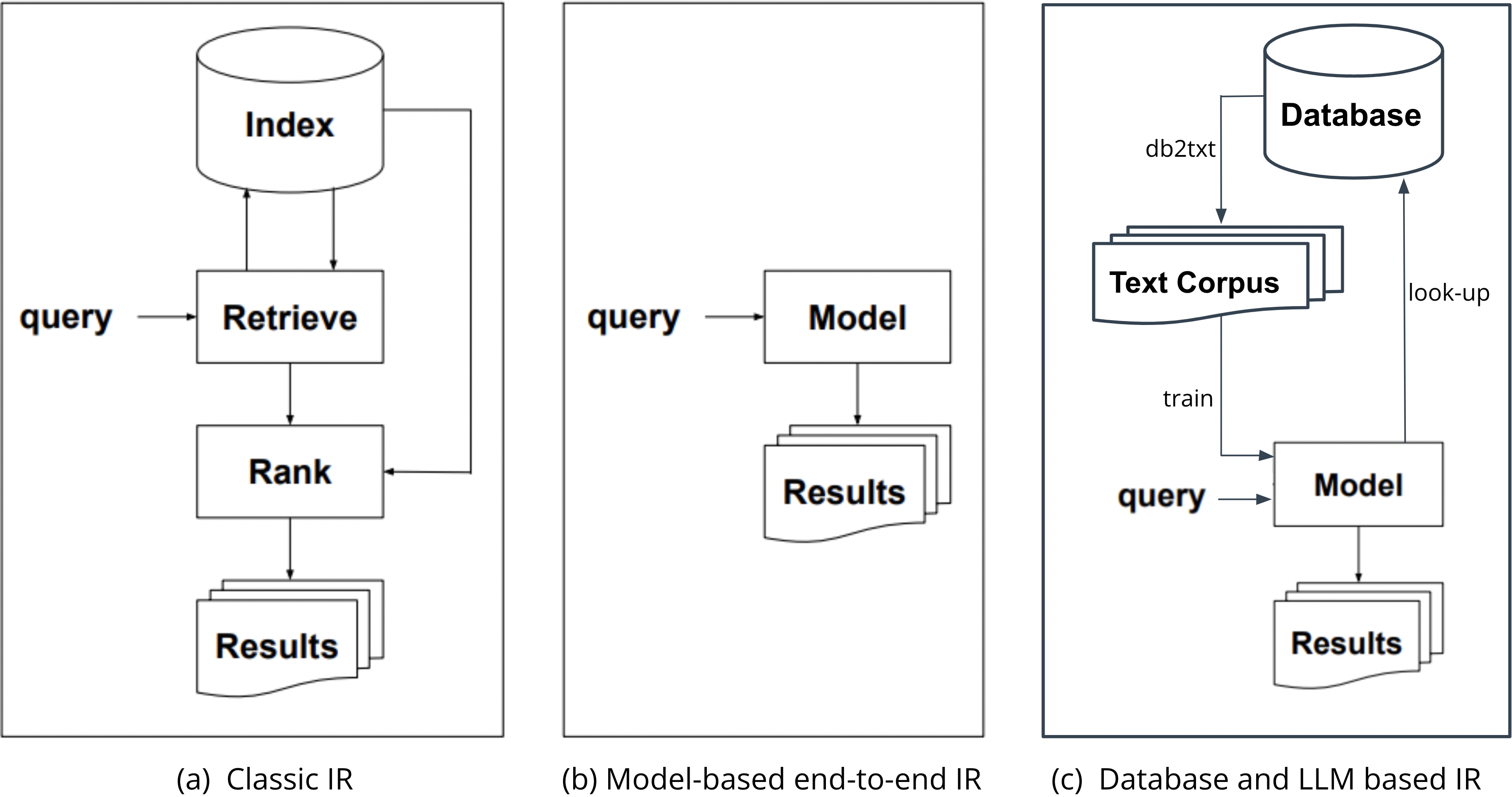}
  \caption{Classic information retrieval (left), model-based information retrieval~\citep{metzler2021rethinking} (middle), and our proposal, a database+LLMs based information retrieval system (right).}
\label{fig:architecture}
\end{figure}

In this section, we describe the architecture of our envisioned solution for e-commerce search. Fig~\ref{fig:architecture} shows the architecture of the classic information retrieval system, the model-based system, and our database and LLM-powered system. 

The classic information retrieval system, as shown in Fig.~\ref{fig:architecture}(a), employs an index-retrieve-then-rank architecture. It requires the development of many for query understanding, retrieval, and ranking, which  results in modern IR systems being comprised of a disparate mix of heterogeneous models (e.g., one model used to learn document representations, another for document understanding, and yet another for ranking)~\citep{metzler2021rethinking}. 
Moreover, the architecture does not support the incorporation of the world knowledge required to understand queries or documents.




\citet{metzler2021rethinking} envisioned a model-based architecture for information retrieval on the web. As shown in Fig. ~\ref{fig:architecture}(b), with a single LLM replacing modules for indexing, retrieval, and ranking, the model-based approach directly provides end-to-end solutions to all kinds of information needs ranging from search to summarization and question answering. The model-based architecture is attractive as it greatly simplifies the IR system, but it faces a number of practical obstacles. 

\begin{enumerate}
    \item An LLM is trained over a text corpus. The model is designed to handle a sequence of tokens instead of structured data, where complex relationships exist among data items.
    \item The model-based approach replaces the traditional index in the IR system. As a result, it does not have knowledge about the universe of document identifiers, and cannot refer to supporting documents in the corpus it is trained over.
    \item Search engines such as Google index billions of new pages every day. It is difficult to continuously update an LLM with such a large amount of data. 
\end{enumerate}

In contrast to web search, IR systems for applications such as e-commerce are centered on structured databases such as product catalogs. On the one hand, a good e-commerce experience for customers requires world knowledge about the products in addition to product specifics in the database, which makes pre-trained LLMs extremely useful as it can provide such knowledge through a unified text-to-text interface. Furthermore, unlike web search engines that must index billions of new pages every day, the structured databases are much limited in size, which are more feasible to be consolidated into or synthesized with LLMs.

On the other hand, the model-based approach is insufficient to support applications such as e-commerce engines. First, because e-commerce search is based on a product catalog, we need a mechanism to consolidate knowledge in the database to the LLM in the training or fine-tuning process. Second, the model must be able to perform database lookups in order to provide product specifics to the customer. In particular, information such as product price and availability that is constantly changing over time is not suitable to be consolidated into the LLM during training or fine-tuning, which makes the ability to refer to records in the database during runtime critical.

Our ultimate goal is to integrate databases and LLMs into a unified infrastructure for information retrieval, with the LLM acting as the single, consolidated model that responds to all types of users' information needs about a set of entities, and the database managing the facts about the entities, including those that are constantly changing.

Specifically, Fig~\ref{fig:architecture}(c) illustrates the architecture behind our vision. In the case of e-commerce search, the database stores products and other types of entities (e.g., brands, manufacturers, etc). We convert the content of the database into textural descriptions, and inside the descriptions, we embed the IDs of the entities. We call such descriptions ``annotated" texts, and we combine them with other text corpora to train an LLM or use them to fine-tune a pre-trained LLM. Essentially, the LLM is trained or fine-tuned with text whose vocabulary has been expanded with IDs. During search, the LLM responds to users' information needs with answers that contain IDs of the relevant entities. Using the IDs, we perform database lookups to retrieve every fact about those entities, including information that changes constantly.

To implement the vision described above, we need to answer three questions.

\begin{enumerate}
    \item How do we create universal IDs for database entities such that they can best connect databases with LLMs?
    \item How do we convert content of the database to annotated texts (descriptions with embedded IDs)?
    \item How do we train or fine-tune LLMs with the annotated texts?
\end{enumerate}

We discuss possible solutions to these questions in Sections 3, 4, and 5, respectively.

\section{Universal IDs}
\label{sec:ids}
To integrate databases and LLMs into a unified IR infrastructure, we add IDs of database records as distinct texts in the text corpus to train or fine-tune LLMs. The LLMs will then respond to user questions with text that contain such IDs, with which we can perform database lookups to retrieve facts about those database records.

The question is how can we best represent the database IDs in  text. The most straightforward way is to represent an ID as is. For example, suppose a record in a product catalog contains the following information: {\tt \{ID:123, name: Milk with Reduced Fat, category: Milk, brand: Lucerne, price: \$10, available: Yes\}}, where {\tt 123} is the ID of the record. We may create a textual description of the product with 123 embedded directly in the description. For example: ``{\tt 123 is a Lucerne brand milk with the name Milk with Reduced Fat.}''

There are two potential problems with the above approach. First, the ID space could be too large. A product catalog may contain millions of items, which means we need to add millions of unique tokens into the vocabulary, whereas the current vocabulary of GPT contains about 50K tokens. Still, this is much better than the web search scenario envisioned by the model-based approach~\citep{metzler2021rethinking}, as Google indexes over 30 billion web pages. A possible solution is to use hierarchical IDs. For e-commerce search, we can leverage the product taxonomy to create hierarchical IDs. In the previous example, instead of {\tt 123}, we may use {\tt Milk/123} as the ID, assuming the product taxonomy has only one level.

The second issue is more subtle. A token such as 123 may already encode some rich knowledge. If we use it as an identifier of a product, the LLM may associate the prior knowledge with the product. A possible solution is to use a rare token that does not have any prior meaning as IDs~\citep{ruiz2022dreambooth}. We can generate a rare token using a random sequence of characters. For example, rather than using the original ID {\tt 123}, we may pick a rare random sequence of characters such as \verb|[K#^&*]| to denote the product, and all we need to do is to maintain a \verb|[K#^&*]| $\leftrightarrow$ {\tt 123} mapping when we need to perform database lookups. 







\section{Unifying All Data into Annotated Texts}
\label{sec:data}
In this section, we describe possible ways of converting structured and semi-structured data to ``annotated text,'' that is, textural description of entities in the (semi-)structured data with embedded IDs. 

Consider a column in a relational table or an edge in a taxonomy or an ontology. Each column/edge denotes either an attribute value or a relationship with another entity. We create a set of templates for every column in every database table and every edge in the taxonomy/ontology. 

\subsection{Manually Created Templates}
For example, consider the product catalog where each record is about a product, and it has columns such as name, brand, manufacturer, etc. We may create the following templates:\\

\noindent{\bf Name (attribute)}:

\begin{itemize}
    \item \verb|[ID]| is \verb|[Name]|.
    \item The name of \verb|[ID]| is \verb|[Name]|.
    \item ...
\end{itemize}

\noindent{\bf Organic (boolean attribute)}:
\begin{itemize}
    \item \verb|[ID]| \verb|[Name]| is (\verb|[Organic]|==true? organic: not organic).
    \item (\verb|[Organic]|==true? \verb|[ID]| is considered healthy as it is organic.)
    \item ...
\end{itemize}
\pagebreak

\noindent{\bf Brand (foreign key)}:

\begin{itemize}
\item The brand ID of  \verb|[ID]| is \verb|[Brand]|.
\item \verb|[ID]| \verb|[Name]| of brand \verb|[Brand]| is manufactured by \verb|[Manufacturer]|
\item ...
\end{itemize}

In the above templates, we use \verb|[ ]| to denote a column in a relational table. Note that each template may refer to multiple columns, as shown in the last template.

For any entity in the relational table, we can instantiate  descriptions using the templates.
For example, for a given product \{{\tt 
 {ID:123, name: Milk with Reduced Fat,
category: Milk, brand: Lucerne, price: \$10, available:
Yes}}\}, we may instantiate the descriptions by following the templates. Here are some results:

\begin{itemize}
\item The name of \verb|[P123]|\footnote{Here we use {\tt [P123]} to denote the token we choose for the entity of ID 123. As described in Section 3, there are multiple options to create such tokens.} is Milk with Reduced Fat.
\item \verb|[P123]| Milk with Reduced Fat is organic.
\item The brand ID of \verb|[P123]| is \verb|[B456]|
\item ...
\end{itemize}

Note that there is no need to instantiate a description using every template for an attribute or a relationship, since they are more or less equivalent, and LLMs have the ability to understand the semantics. The reason to have multiple templates and randomly sample from them is to avoid bias.

\subsection{Query Based Templates}

The templates described above are designed to describe an entity one attribute or one relationship a time, but a relational table contains important information and knowledge that only manifests if we examine the data in a more holistic way.

We propose to create templates based on database queries. Each query has a meaning behind it, and through templates, we will express such  meanings in natural language. To a certain extent, this could be considered the reverse of the Text-to-SQL~\citep{katsogiannis2023survey} effort. 

We are particularly interested in database queries that contain joins and aggregates, which enable us to express more complex information. For example, joins help us connect grocery ingredients and recipes, and products and complementary products, etc.
As an example, we can generate descriptions such as ``A glass of \verb|[P103]| Reduced fat Lucerne Milk contains about 140 calories,'' which comes from the result when we join the product table with a nutrition table (and calculate calories based on {\tt servingSize} and {\tt fatCaloriesPerServing}).
 
The user engagement data is critical to e-commerce search. To generate natural language descriptions from user engagement data, we will use templates based on aggregate and group-by queries. This allows us to convey in natural language information such as ``how frequently a product is purchased'', ``which products customers purchase together'', etc. The resulting annotated text may look like these:

\begin{itemize}
\item The top 5 items customers purchase after searching for `milk' at Instacart are \verb|[P101]|, \verb|[P102]|, ..., and \verb|[P105]|.
\item Over 1,000 customers buy \verb|[P001]|, \verb|[P022]|, and \verb|[P042]| together every day in California.
\item ...
\end{itemize}


\begin{figure}[!b]
\centering
\begin{tabular}{|p{0.15\textwidth}|p{0.8\textwidth}|}
\hline
\textbf{Prompt} & 

Could you use five different ways to describe the following product in English? Please make it sound natural, and please embed the id in each description.

\{

\quad ``id": {\tt [P123]},
 
\quad ``name": ``Lucerne Milk with Reduced Fat",
 
\quad ``description": ``the best milk in the world",
 
\quad ``icon": ``http://instacart.com/img/p103.png",
 
\quad ``brand": ``Lucerne",
 
\quad ``feature": ``organic, gluten free, and kosher",
 
\quad ``price (\$)": 3.00
 
\}
\\ \hline
\textbf{LLM \newline Response} & 






\parbox{\linewidth}{
1. Check out the Lucerne Milk with Reduced Fat, product ID {\tt [P123]}. It's touted as the best milk in the world, and it's organic, gluten free, and kosher. Plus, it's only \$3.00! You can see what it looks like on the Instacart website. \newline

2. For just \$3.00, you can enjoy the world's best milk, Lucerne Milk with Reduced Fat. It's organic, gluten free, and kosher. You can find it under the product ID {\tt [P123]} on our website, where you can also view its image. \newline

3. Have you tried the Lucerne Milk with Reduced Fat? It's product ID {\tt [P123]} and it's known as the best milk in the world. It's organic, gluten free, and kosher, and it's only \$3.00. You can see its image on our website. \newline

4. The Lucerne Milk with Reduced Fat, product ID {\tt [P123]}, is not just any milk. It's the best milk in the world, and it's organic, gluten free, and kosher. Plus, it's only \$3.00. Check out its image on our website. \newline

5. If you're looking for the best milk in the world, try the Lucerne Milk with Reduced Fat. It's organic, gluten free, and kosher, and it's only \$3.00. You can find it on our website under the product ID {\tt [P123]}.}

\\ \hline
\end{tabular}
\caption{Using LLMs such as ChatGPT to automatically generate descriptions of database records in natural language.}
\label{fig:gpt3}
\end{figure}
\FloatBarrier
 
\subsection{LLM Generated Templates}

We have described how to manually create templates that can be instantiated in natural language. If the database contains a large number of tables with a complex schema, manually creating templates may be a time-consuming process, and the resulting templates may lack diversity.

We can also use LLMs such as ChatGPT to automatically generate templates. Fig.~\ref{fig:gpt3} shows an example, where ChatGPT is asked to generate a set of descriptions for a database record and the descriptions must embed the ID of the record. As we can see the diversity of the descriptions is quite high. The idea is not to invoke ChatGPT to describe every record in a database table. Rather, we can convert the descriptions that ChatGPT generates for a few records to templates and then use the templates to generate natural language descriptions of each record.

We may also use ChatGPT to generate query-based templates. However, directly using SQL as input to ChatGPT hasn't produced good results. Instead, we can ask ChatGPT to describe the results of the SQL queries. In other words, we convert this problem to the problem in Fig.~\ref{fig:gpt3}

\begin{figure*}[t!]
\centering
\includegraphics[width=1.0\textwidth]{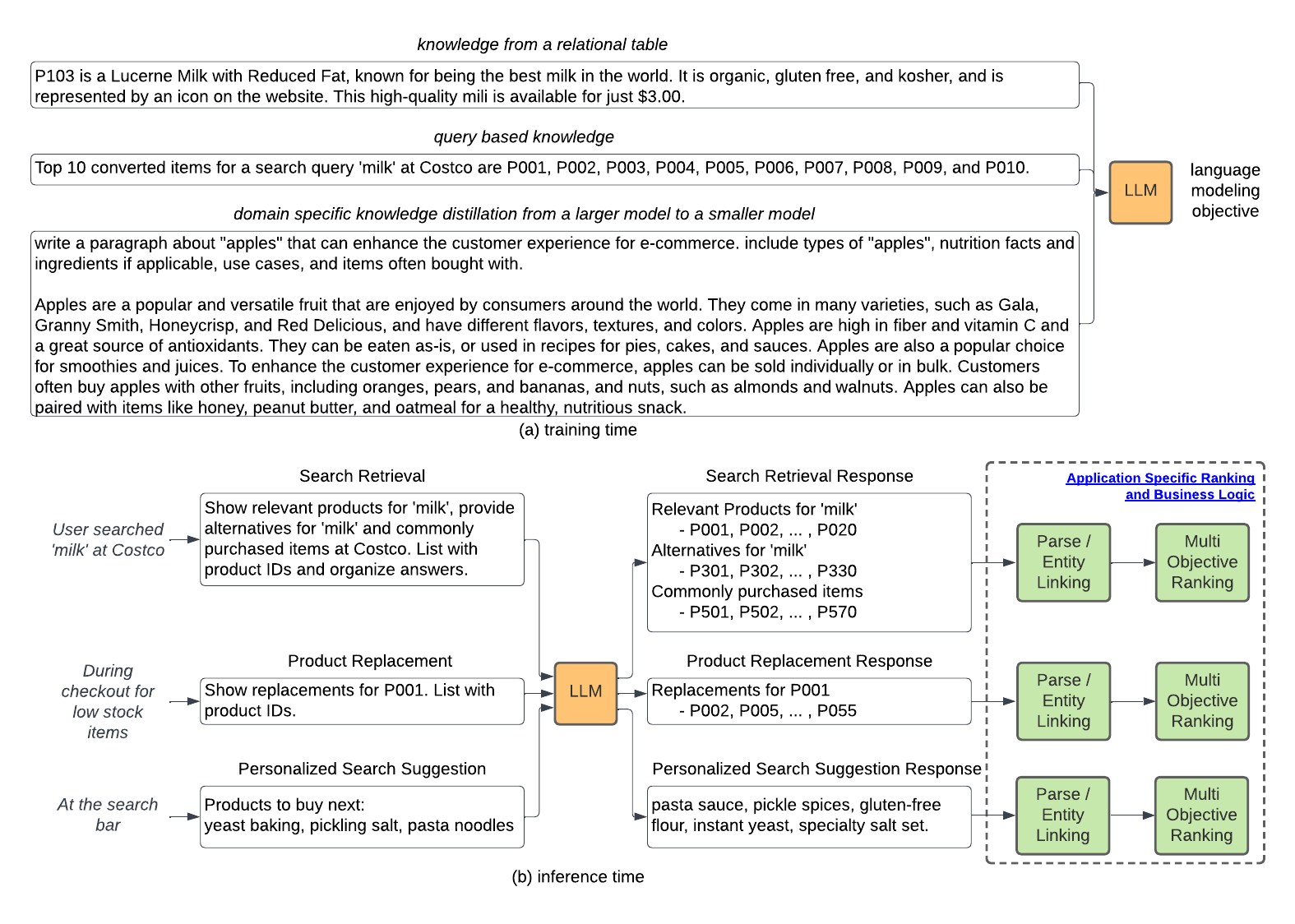}
\caption{Example of how an LLM can be used during training and inference time.
(a) shows training time where an LLM learns knowledge generated from a relational database and database queries. Optionally, the model can learn richer domain-specific world knowledge through distillation.
(b) shows inference time behavior. The example on the top shows a model that performs a search retrieval, where the model shows relevant, alternative, and commonly purchased items as product IDs. Retrieved product IDs are then processed for final multi-objective ranking and business logic.
The example in the middle shows that the model recommends replacements for a specific product ID.
The last example shows that the model recommends next search suggestions given a search history.}
\label{fig:system}
\end{figure*}

\section{System Architecture}
\label{sec:system}
Once we convert all the structured data into non-structured data as natural language text, we ingest it into LLM with a traditional language modeling approach.
During training, our goal is to transfer the knowledge learned from our database to a LLM while maintaining semantic and syntactic properties of language in the LLM.
Once fine-tuning is complete, the model becomes proficient in using domain-specific e-commerce world knowledge in conjunction with information obtained from a database.
This enables us to utilize the model as a retrieval and recommendation engine during inference, with product IDs seamlessly integrated into the model itself.

\subsection{Training}
\label{sec:training}
We envision that the model-based search can be realized by fine-tuning an LLM with a language modeling (LM) objective using the annotated texts converted from structured data. Multi-task learning using specific task identifiers~\citep{metzler2021rethinking, t5} can also be utilized to train a model, but fine-tuning an LLM with an LM objective using texts has several benefits.

\begin{itemize}
\item It doesn’t require any labeled data, whereas the previous approach~\citep{metzler2021rethinking} requires some supervision. While labeled data (e.g., positive relationships between a query and products) can be utilized, it is not employed for supervised learning in our case.
Instead, we first transform labeled data into natural language text format, as outlined in Section ~\ref{sec:data}, and then train a model with self-supervised learning.
\item The use of LM objectives is a highly effective method for transferring knowledge, mirroring the objectives employed in pre-trained LLM training.
\end{itemize}

Figure ~\ref{fig:system} (a) depicts the process of training a LM. We start with a pre-trained LLM and fine-tune the model with the annotated texts obtained from the methods described in Section \ref{sec:data}.
We optionally inject general world knowledge useful for the e-commerce domain as additional examples to encourage the model to retain world knowledge specific to the e-commerce domain and linguistic syntax and semantics of the language in the LLM.
This step allows us to naturally transfer domain knowledge to the LLM through an innate ability to learn and memorize.
During training, the model memorizes product IDs as novel vocabularies while also establishing correlations between these new concepts and the pre-existing world knowledge embedded in the model.


\subsection{Inference}
\label{sec:inference}
Once we train the model, we utilize the model as a retrieval and recommendation engine to perform various tasks as shown in figure ~\ref{fig:system} (b).
Similar to conventional prompt engineering, we can leverage the model with various configurations such as zero-shot, few-shot, and fine-tuning as in ~\citep{gpt3}.
The examples in figure ~\ref{fig:system} (b) show how to use the model in a zero-shot setting.
We show search retrieval, product replacement, and personalized search suggestions as example use cases.
To generate answers with specific product IDs, we need to be clear and precise with the LLM.
We list several prompts below.

\begin{itemize}
\item \textbf{search retrieval}: Show relevant products for 'milk', provide alternatives for 'milk' and commonly purchased items at Costco. List with product IDs and organize answers.
\item \textbf{product replacement}: Show replacements for {\tt [P001]}. List with product IDs.
\item \textbf{personalized search suggestion}: Products to buy next: yeast baking, pickling salt, pasta noodles.
\end{itemize}

By using the prompts outlined above, we expect that the LLM will generate responses in a pre-defined manner.
Upon generating the output, as illustrated in Figure ~\ref{fig:system} (b), we extract information using simple text manipulation techniques or an additional entity linking model.
Instructing the LLM to output in JSON format is also a common practice to enforce desired formatting.
Following this, we obtain the most up-to-date information, including pricing and availability, by connecting to a real-time updated database. Additional ranking and business logic can be applied, and the final results are presented to users.

\section{Research Directions}
\label{sec:research}
There are numerous obstacles that need to be overcome in order to realize our vision in a production environment. We shall delve into several issues that are worth exploring in the subsequent sections.

\subsection{Latency Optimization}
\label{sec:optimization}

The primary obstacle to implementing model-based search in a production environment is latency.
Since language models generate output responses token by token, the latency of the output is heavily influenced by the length of the response being generated.
We explore several approaches to overcoming this challenge in the following section.

\subsubsection{{The Encoder Based Approach}}
An encoder based model with fine tuned classification heads can generate output with a single inference.
In this approach, the model generates a probability score for each class (a product in e-commerce) simultaneously.

\begin{figure*}[t!]
\centering
\includegraphics[width=1\textwidth]{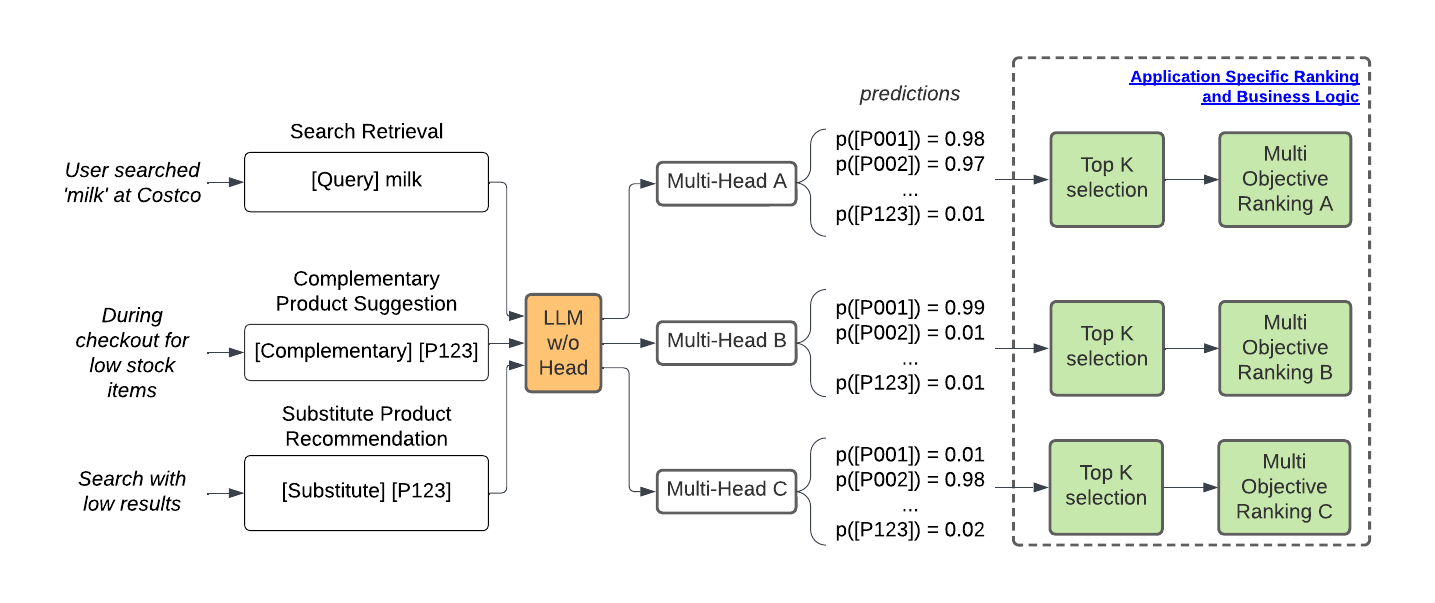}
\caption{Example of how a single model can be used for multiple tasks with an encoder model approach. This example shows a model that performs a search relevance task, complementary product suggestion, and substitute product recommendation.}
\label{fig:encoder-continuous-update}
\end{figure*}

Using the LLM trained with unstructured data integrated with the product IDs as described in section ~\ref{sec:training}, the model can be further fine-tuned with supervised tasks.
The last projection layer that was used for token generation is removed, and an aggregation layer is added, often incorporating pooling methods, to produce refined representations. Finally, a head with an output size that matches the number of classes (products) is added to create a classification model, as depicted in Figure ~\ref{fig:encoder-continuous-update}.



Figure ~\ref{fig:encoder-continuous-update} shows several tasks that can be performed with an encoder based approach. We show the T5~\citep{t5}  model approach as an example where task identifiers can be used to perform multiple tasks using a single model. For the search relevance task, we treat this problem as a multiclass-multilabel classification task, where \# of classes = \# of product identifiers. For a single query, we can map multiple relevant product IDs, and can feed into the model. A sample training example is shown in the following.

\begin{itemize}
\item “{\tt [Query]} milk”: ({\tt [P001]}, 0.4), ({\tt [P006]}, 0.3), ({\tt [P123]}, 0.2), … , ({\tt [P234]}, 0.1)
\end{itemize}

“{\tt [Query]}” indicates the task is a search relevance task, where labels can be soft or hard (we can utilize previous user engagement history as soft labels).

Another task can be complementary product suggestions. Given a product, we predict multiple complementary products that can go well together. This problem can also be solved as a multiclass-multilabel classification task, and below is a sample training example.

\begin{itemize}
\item “{\tt [Complementary]} {\tt [P001]}”: ({\tt [P006]}, 0.8), ({\tt [P123]}, 0.9), … , ({\tt [P234]}, 0.7)
\end{itemize}

Here, ``{\tt [Complementary]}" represents the task as a complementary product suggestion.

Next task shown in the figure is substitute product recommendation where it is very similar to complementary product suggestion task in terms of input and output types but with a different task identifier. The task identifier used for this task is “{\tt [Substitute]}”.

Once we fine-tune the model, the inference for the encoder model approach becomes faster compared to the generative language model. We simply use a desired task identifier with a query (a search term or a product ID text), perform a single inference of the model, and obtain the final probability scores for all product IDs from the model.
The parsing and entity linking used for the generative model approach can now be replaced with top K selection logic to get the final candidate product IDs for each task.
Based on the requirements of each domain, we perform the final ranking and present the results to the users.

The benefit of using an encoder-based approach is that it has predictable inference time complexity since it can directly produce product IDs as results using a single inference. Single inference per task is advantageous, leading to lower run-time costs. However, this approach has several drawbacks.
\begin{itemize}
\item A supervised learning dataset is necessary for each task in a multi-task learning process. This can pose a significant challenge, especially for cold start products.
\item The number of classes in the output layer increases as more product IDs are added, which can result in practical challenges during implementation.
\item It is not scalable since each classification model has to be trained with new training objectives whenever we need a different task.
\end{itemize}

One can utilize both the benefits of generative language model and classification model. For the most popular tasks, we can utilize the encoder model approach for faster inference while we can quickly prototype new applications with the generative approach.

Most production machine learning models are based on the encoder model, as it still necessitates strict latency requirements for deployment in production. However, with the continuous improvement of computing resources, the latency of the generative language model is also improving. It will be intriguing to see how this develops in the future.

\subsubsection{{Model Compression:}}
Another research direction to optimize the latency of the models is model compression.
At a high level, model compression aims to optimize inference time latency either by (1) knowledge distillation ~\citep{gu2023knowledge, agarwal2023gkd, jha2023large}, which involves training a smaller model with the rich knowledge found in a larger model, (2) pruning ~\citep{ma2023llmpruner, zhang2023pruning}, which involves removing redundant computations, or (3) quantization ~\citep{liu2023llmqat, kim2023memoryefficient, kim2023squeezellm, lin2023awq}, which involves approximating computation.
Given the intense competition to host in-house LLMs using cost-effective methods, it will be interesting to observe the progression of these lines of research.


\subsection{Personalization}
\label{sec:personalization}

\begin{figure*}[t!]
\centering
\includegraphics[width=1.0\textwidth]{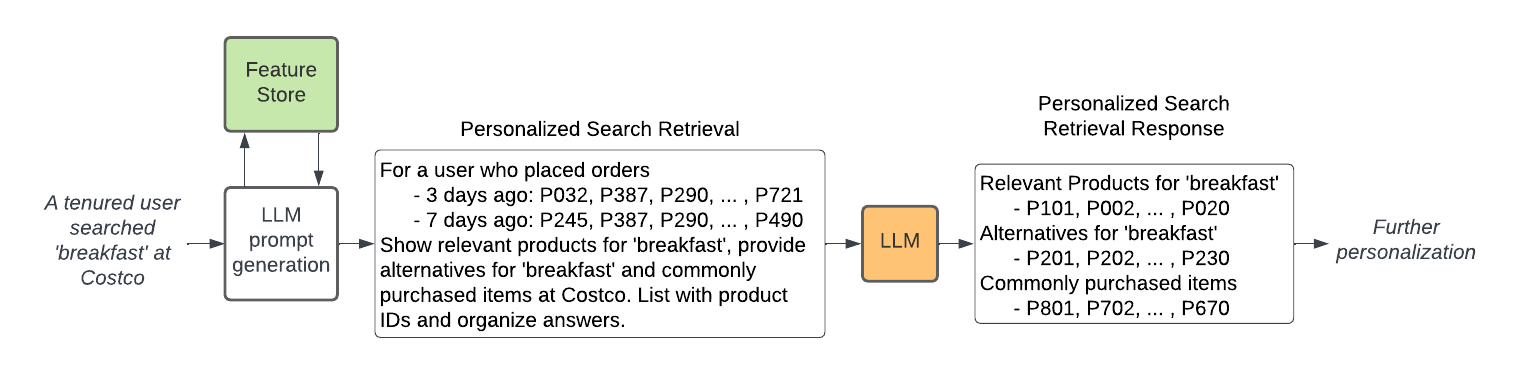}
\caption{The above example shows how to personalize search by incorporating user engagement history in the prompt for a LLM. When a user issues a query, user engagement signals like past orders can be retrieved from a feature store, then those are provided as a context in the input prompt.
This is especially useful for broad intent queries like 'breakfast' when there are many relevant items, it helps retrieve better personalized items.
After the LLM response, further personalization can be applied using another ML models. Other applications can also follow similar prompt engineering to personalize the experience.}
\label{fig:personalization}
\end{figure*}

Personalization has been a crucial subject in e-commerce applications, and many related studies have employed sequence modeling to enhance the comprehension of user intent in the search or recommendation arena ~\citep{li2021taobao, Pancha2022PinnerFormerSM}. We anticipate that all of these approaches can be applied concurrently with LLM, enabling the utilization of world knowledge in LLM and sophisticated user modeling.

As demonstrated in Figure ~\ref{fig:personalization}, we can leverage LLM to personalize search. The concept involves incorporating past user engagement data as input to the prompt and allowing LLM to generate a response. The feature store, as depicted in the figure, comprises user-level features such as prior purchases, which are integrated into the prompt.
Once the LLM generates a response, it can be used as input for further personalization. 
A conventional user sequence model can be employed to deliver accurate predictions. The example in the figure illustrates a zero-shot configuration, but we can adopt few-shot learning and/or fine-tuning approaches to achieve more sophisticated outcomes.




\subsection{Catastrophic Forgetting}
\label{sec:catastrophic-forgetting}
Product information in e-commerce keeps getting updated, we need to continue to ingest the updated and new data into the model. Whenever the new information needs to be ingested, we can perform fine-tuning the model with the new data.
As the size of the database grows, however, we might encounter a catastrophic forgetting problem~\citep{FRENCH1999catastrophic} where we completely forget the old data that has been used for training the model.
In order to ensure we don’t encounter catastrophic forgetting problems, we might want to use a big enough model in order for the model to contain all the information. However, the absence of a clear guideline makes it difficult to determine the right model capacity based on the size of the data.
A number of methods~\citep{DeLange2019ACL, ahn2019uncertainty, Nguyen2019TowardUC, Serr2018OvercomingCF} have been proposed to address this issue, but most of the research has focused on small or simulated environments.

An alternative approach to managing this is to restrict the volume of data we need to ingest. Rather than considering product IDs as separate documents, we can use internal taxonomies as unique documents for broad information intake. These taxonomies are typically much smaller in size compared to individual products. 
By treating the search and recommendation task at the taxonomy level, the amount of information the LLM needs to learn will be considerably reduced, thereby addressing the catastrophic forgetting issue.
The study of catastrophic forgetting in LLMs continues to be a fascinating and demanding area of research.


\section{Related Work}
\label{sec:related}
Our proposed approaches try to incorporate external (e-commerce) data into LLM as data storage. Since the model understands embedded knowledge, it can be directly used to generate output responses without external help during inference time. The runtime convenience comes at the expense of training the model.

As discussed in earlier Section~\ref{sec:research}, continuous update with training can lead to problems like catastrophic forgetting and language drift.
In order to address these problems, there have been several efforts~\citep{Liu2022gptindex, chase2022langchain, Guu2020realm, lewis2020rag, nakano2021webgpt} to decouple external data ingestion and output generation without training the model.
The idea here is to use LLM to generate output responses using the external data as context.
This is referred to as retrieval augmented generation (RAG) ~\citep{lewis2020rag}.
They try to overcome the limitations of a fixed context window used in LLM.
Since we cannot give our entire database as input prompt, all the approaches involve fetching the data, augmenting and refining it.
Since refinement is a lossy process, it is possible that the input prompt may not capture all the essential context which is vital for output generation.
This can be a critical issue especially for e-commerce applications where there are many relevant contexts (products) for any given task.
When there are many documents in need of refinement, additional LLM calls for refinement can cause latency increase.
Although there are limitations in the above approach, it will be interesting to see the progress of research trying to decouple data ingestion and output generation process.

Text-to-SQL parsing~\citep{katsogiannis2023survey} is an area of research that tries to transform natural language questions into their corresponding SQL statements, based on data provided by relational databases. Conceptually this is similar to what the current e-commerce engine is doing, but it has limitations. The tasks that can be performed with SQL statements are confined by SQL syntax and database schema, it becomes difficult to perform queries that require common knowledge (e.g. what are relevant product IDs that are good for Christmas?). SQL cannot generate meaningful responses without having explicit knowledge embedded in databases. Keeping the explicit knowledge as a knowledge graph in the database can be a solution, but it is not scalable. Use of knowledge in LLM right from the retrieval of documents can make complex tasks achievable.

\section{Conclusion}
\label{sec:conclusion}
The two biggest challenges for e-commerce search are understanding queries and leveraging world knowledge to match queries and products. 
Large language models are not only powerful in query understanding, but they also possess a vast amount of world knowledge. Still, databases are a critical component of the e-commerce infrastructure as they manage accurate and sometimes time-changing facts about the products. This position paper examines how the strengths of database systems and large language models can be synthesized to create information retrieval systems that better support e-commerce search. We believe the solution is to convert structured and semi-structured data such as the product catalog, taxonomies, and ontology to natural language text and train a language model, which is used as an end-to-end solution for e-commerce search. This is a clear departure from previous e-commerce search approaches that focus on converting unstructured data such as product descriptions, customer reviews, web pages, etc. to structured data through a costly information extraction process and using a myriad of algorithms and machine learning models to support a complex system with separate modules for indexing, retrieval, and ranking.

\bibliography{sample-base}

\begin{thebibliography}{29}
\providecommand{\natexlab}[1]{#1}
\providecommand{\url}[1]{\texttt{#1}}
\expandafter\ifx\csname urlstyle\endcsname\relax
  \providecommand{\doi}[1]{doi: #1}\else
  \providecommand{\doi}{doi: \begingroup \urlstyle{rm}\Url}\fi

\bibitem[Agarwal et~al.(2023)Agarwal, Vieillard, Stanczyk, Ramos, Geist, and Bachem]{agarwal2023gkd}
Rishabh Agarwal, Nino Vieillard, Piotr Stanczyk, Sabela Ramos, Matthieu Geist, and Olivier Bachem.
\newblock Gkd: Generalized knowledge distillation for auto-regressive sequence models, 2023.

\bibitem[Ahn et~al.(2019)Ahn, Cha, Lee, and Moon]{ahn2019uncertainty}
Hongjoon Ahn, Sungmin Cha, Donggyu Lee, and Taesup Moon.
\newblock Uncertainty-based continual learning with adaptive regularization.
\newblock In \emph{Advances in Neural Information Processing Systems}, pages 4394--4404, 2019.

\bibitem[Brown et~al.(2020)Brown, Mann, Ryder, Subbiah, Kaplan, Dhariwal, Neelakantan, Shyam, Sastry, Askell, et~al.]{gpt3}
Tom Brown, Benjamin Mann, Nick Ryder, Melanie Subbiah, Jared~D Kaplan, Prafulla Dhariwal, Arvind Neelakantan, Pranav Shyam, Girish Sastry, Amanda Askell, et~al.
\newblock Language models are few-shot learners.
\newblock \emph{Advances in neural information processing systems}, 33:\penalty0 1877--1901, 2020.

\bibitem[Chase(2022)]{chase2022langchain}
Harrison Chase.
\newblock {LangChain}, 11 2022.
\newblock URL \url{https://github.com/hwchase17/langchain}.

\bibitem[French(1999)]{FRENCH1999catastrophic}
Robert~M. French.
\newblock Catastrophic forgetting in connectionist networks.
\newblock \emph{Trends in Cognitive Sciences}, 3\penalty0 (4):\penalty0 128--135, 1999.
\newblock ISSN 1364-6613.
\newblock \doi{https://doi.org/10.1016/S1364-6613(99)01294-2}.
\newblock URL \url{https://www.sciencedirect.com/science/article/pii/S1364661399012942}.

\bibitem[Gu et~al.(2023)Gu, Dong, Wei, and Huang]{gu2023knowledge}
Yuxian Gu, Li~Dong, Furu Wei, and Minlie Huang.
\newblock Knowledge distillation of large language models, 2023.

\bibitem[Guu et~al.(2020)Guu, Lee, Tung, Pasupat, and Chang]{Guu2020realm}
Kelvin Guu, Kenton Lee, Zora Tung, Panupong Pasupat, and Ming-Wei Chang.
\newblock Realm: Retrieval-augmented language model pre-training.
\newblock In \emph{Proceedings of the 37th International Conference on Machine Learning}, ICML'20. JMLR.org, 2020.

\bibitem[Huang et~al.(2020)Huang, Sharma, Sun, Xia, Zhang, Pronin, Padmanabhan, Ottaviano, and Yang]{huang2020embedding}
Jui-Ting Huang, Ashish Sharma, Shuying Sun, Li~Xia, David Zhang, Philip Pronin, Janani Padmanabhan, Giuseppe Ottaviano, and Linjun Yang.
\newblock Embedding-based retrieval in facebook search.
\newblock In \emph{Proceedings of the 26th ACM SIGKDD International Conference on Knowledge Discovery \& Data Mining}, pages 2553--2561, 2020.

\bibitem[Jha et~al.(2023)Jha, Groeneveld, Strubell, and Beltagy]{jha2023large}
Ananya~Harsh Jha, Dirk Groeneveld, Emma Strubell, and Iz~Beltagy.
\newblock Large language model distillation doesn't need a teacher, 2023.

\bibitem[Katsogiannis-Meimarakis and Koutrika(2023)]{katsogiannis2023survey}
George Katsogiannis-Meimarakis and Georgia Koutrika.
\newblock A survey on deep learning approaches for text-to-sql.
\newblock \emph{The VLDB Journal}, pages 1--32, 2023.

\bibitem[Kim et~al.(2023{\natexlab{a}})Kim, Lee, Kim, Park, Yoo, Kwon, and Lee]{kim2023memoryefficient}
Jeonghoon Kim, Jung~Hyun Lee, Sungdong Kim, Joonsuk Park, Kang~Min Yoo, Se~Jung Kwon, and Dongsoo Lee.
\newblock Memory-efficient fine-tuning of compressed large language models via sub-4-bit integer quantization, 2023{\natexlab{a}}.

\bibitem[Kim et~al.(2023{\natexlab{b}})Kim, Hooper, Gholami, Dong, Li, Shen, Mahoney, and Keutzer]{kim2023squeezellm}
Sehoon Kim, Coleman Hooper, Amir Gholami, Zhen Dong, Xiuyu Li, Sheng Shen, Michael~W. Mahoney, and Kurt Keutzer.
\newblock Squeezellm: Dense-and-sparse quantization, 2023{\natexlab{b}}.

\bibitem[Lange et~al.(2019)Lange, Aljundi, Masana, Parisot, Jia, Leonardis, Slabaugh, and Tuytelaars]{DeLange2019ACL}
Matthias~De Lange, Rahaf Aljundi, Marc Masana, Sarah Parisot, Xu~Jia, Ales Leonardis, Gregory~G. Slabaugh, and Tinne Tuytelaars.
\newblock A continual learning survey: Defying forgetting in classification tasks.
\newblock \emph{IEEE Transactions on Pattern Analysis and Machine Intelligence}, 44:\penalty0 3366--3385, 2019.

\bibitem[Lewis et~al.(2020)Lewis, Perez, Piktus, Petroni, Karpukhin, Goyal, K\"{u}ttler, Lewis, Yih, Rockt\"{a}schel, Riedel, and Kiela]{lewis2020rag}
Patrick Lewis, Ethan Perez, Aleksandra Piktus, Fabio Petroni, Vladimir Karpukhin, Naman Goyal, Heinrich K\"{u}ttler, Mike Lewis, Wen-tau Yih, Tim Rockt\"{a}schel, Sebastian Riedel, and Douwe Kiela.
\newblock Retrieval-augmented generation for knowledge-intensive nlp tasks.
\newblock In \emph{Advances in Neural Information Processing Systems}, volume~33, pages 9459--9474. Curran Associates, Inc., 2020.

\bibitem[Li et~al.(2021)Li, Lv, Jin, Lin, Yang, Zeng, Wu, and Ma]{li2021taobao}
Sen Li, Fuyu Lv, Taiwei Jin, Guli Lin, Keping Yang, Xiaoyi Zeng, Xiao-Ming Wu, and Qianli Ma.
\newblock Embedding-based product retrieval in taobao search.
\newblock In \emph{Proceedings of the 27th ACM SIGKDD}, KDD '21, page 3181–3189, New York, NY, USA, 2021. ACM.
\newblock ISBN 9781450383325.

\bibitem[Lin et~al.(2023)Lin, Tang, Tang, Yang, Dang, and Han]{lin2023awq}
Ji~Lin, Jiaming Tang, Haotian Tang, Shang Yang, Xingyu Dang, and Song Han.
\newblock Awq: Activation-aware weight quantization for llm compression and acceleration.
\newblock \emph{arXiv}, 2023.

\bibitem[Liu(2022)]{Liu2022gptindex}
Jerry Liu.
\newblock {GPT Index}, 11 2022.
\newblock URL \url{https://github.com/jerryjliu/gpt_index}.

\bibitem[Liu et~al.(2023)Liu, Oguz, Zhao, Chang, Stock, Mehdad, Shi, Krishnamoorthi, and Chandra]{liu2023llmqat}
Zechun Liu, Barlas Oguz, Changsheng Zhao, Ernie Chang, Pierre Stock, Yashar Mehdad, Yangyang Shi, Raghuraman Krishnamoorthi, and Vikas Chandra.
\newblock Llm-qat: Data-free quantization aware training for large language models, 2023.

\bibitem[Ma et~al.(2023)Ma, Fang, and Wang]{ma2023llmpruner}
Xinyin Ma, Gongfan Fang, and Xinchao Wang.
\newblock Llm-pruner: On the structural pruning of large language models.
\newblock In \emph{Advances in Neural Information Processing Systems}, 2023.

\bibitem[Metzler et~al.(2021)Metzler, Tay, Bahri, and Najork]{metzler2021rethinking}
Donald Metzler, Yi~Tay, Dara Bahri, and Marc Najork.
\newblock Rethinking search: making domain experts out of dilettantes.
\newblock In \emph{ACM SIGIR Forum}, volume~55, pages 1--27. ACM New York, NY, USA, 2021.

\bibitem[Nakano et~al.(2021)Nakano, Hilton, Balaji, Wu, Ouyang, Kim, Hesse, Jain, Kosaraju, Saunders, Jiang, Cobbe, Eloundou, Krueger, Button, Knight, Chess, and Schulman]{nakano2021webgpt}
Reiichiro Nakano, Jacob Hilton, Suchir Balaji, Jeff Wu, Long Ouyang, Christina Kim, Christopher Hesse, Shantanu Jain, Vineet Kosaraju, William Saunders, Xu~Jiang, Karl Cobbe, Tyna Eloundou, Gretchen Krueger, Kevin Button, Matthew Knight, Benjamin Chess, and John Schulman.
\newblock Webgpt: Browser-assisted question-answering with human feedback, 2021.
\newblock URL \url{https://arxiv.org/abs/2112.09332}.

\bibitem[Nguyen et~al.(2019)Nguyen, Achille, Lam, Hassner, Mahadevan, and Soatto]{Nguyen2019TowardUC}
Cuong~V Nguyen, Alessandro Achille, Michael Lam, Tal Hassner, Vijay Mahadevan, and Stefano Soatto.
\newblock Toward understanding catastrophic forgetting in continual learning.
\newblock \emph{ArXiv}, abs/1908.01091, 2019.

\bibitem[Pancha et~al.(2022)Pancha, Zhai, Leskovec, and Rosenberg]{Pancha2022PinnerFormerSM}
Nikil Pancha, Andrew Zhai, Jure Leskovec, and Charles~R. Rosenberg.
\newblock Pinnerformer: Sequence modeling for user representation at pinterest.
\newblock \emph{Proceedings of the 28th ACM SIGKDD Conference on Knowledge Discovery and Data Mining}, 2022.

\bibitem[Raffel et~al.(2020)Raffel, Shazeer, Roberts, Lee, Narang, Matena, Zhou, Li, and Liu]{t5}
Colin Raffel, Noam Shazeer, Adam Roberts, Katherine Lee, Sharan Narang, Michael Matena, Yanqi Zhou, Wei Li, and Peter~J Liu.
\newblock Exploring the limits of transfer learning with a unified text-to-text transformer.
\newblock \emph{The Journal of Machine Learning Research}, 21\penalty0 (1):\penalty0 5485--5551, 2020.

\bibitem[Ruiz et~al.(2022)Ruiz, Li, Jampani, Pritch, Rubinstein, and Aberman]{ruiz2022dreambooth}
Nataniel Ruiz, Yuanzhen Li, Varun Jampani, Yael Pritch, Michael Rubinstein, and Kfir Aberman.
\newblock Dreambooth: Fine tuning text-to-image diffusion models for subject-driven generation.
\newblock 2022.

\bibitem[Serr{\`a} et~al.(2018)Serr{\`a}, Sur{\'i}s, Miron, and Karatzoglou]{Serr2018OvercomingCF}
Joan Serr{\`a}, D{\'i}dac Sur{\'i}s, Marius Miron, and Alexandros Karatzoglou.
\newblock Overcoming catastrophic forgetting with hard attention to the task.
\newblock In \emph{International Conference on Machine Learning}, 2018.

\bibitem[Xie et~al.(2022)Xie, Na, Xiao, Manchanda, Rao, Xu, Shu, Vasiete, Tenneti, and Wang]{na2022embedding}
Yuqing Xie, Taesik Na, Xiao Xiao, Saurav Manchanda, Young Rao, Zhihong Xu, Guanghua Shu, Esther Vasiete, Tejaswi Tenneti, and Haixun Wang.
\newblock An embedding-based grocery search model at instacart.
\newblock \emph{SIGIR 2022 Workshop on eCommerce. ACM.}, 2022.

\bibitem[Zhang et~al.(2020)Zhang, Wang, Zhang, Tang, Jiang, Xiao, Yan, and Yang]{zhang2020retrieval}
Han Zhang, Songlin Wang, Kang Zhang, Zhiling Tang, Yunjiang Jiang, Yun Xiao, Weipeng Yan, and Wen-Yun Yang.
\newblock Towards personalized and semantic retrieval: An end-to-end solution for e-commerce search via embedding learning.
\newblock In \emph{Proceedings of the 43rd International ACM SIGIR Conference on Research and Development in Information Retrieval}, pages 2407--2416, 2020.

\bibitem[Zhang et~al.(2023)Zhang, Chen, Shen, Yang, Ou, Yu, and Zhuang]{zhang2023pruning}
Mingyang Zhang, Hao Chen, Chunhua Shen, Zhen Yang, Linlin Ou, Xinyi Yu, and Bohan Zhuang.
\newblock Pruning meets low-rank parameter-efficient fine-tuning, 2023.

\end{thebibliography}
\end{document}